\documentclass[aps,prd,amsfonts,superscriptaddress,nofootinbib,eqsecnum,
              secnumarabic,tightenlines,showpacs]{revtex4}
\usepackage{bm,hyperref,mathrsfs}
\newcommand{\getp}{{\get^\parallel}}
\newcommand{\getn}{{\get^\perp}}

\newcommand{\vc}[1]{{\bm{#1}}}
\newcommand{\der}{\partial}

\DeclareMathSymbol{\mg}{\mathrel}{symbols}{"1D} 

\newcommand{\mx}[1]{\bm{#1}}

\newcommand{\Bnabla}{\bm{\nabla}}

\newcommand{\half}{\frac{1}{2}}
\newcommand{\Id}{\openone}

\newcommand{\ra}{\rightarrow}
\newcommand{\inv}{^{-1}}
\newcommand{\lh}{\left(}
\newcommand{\rh}{\right)}
\newcommand{\labl}[1]{\label{#1}}
\newcommand{\eqref}[1]{(\ref{#1})}

\renewenvironment{matrix}{%
  \hskip -\arraycolsep\array{cccccccccc}%
}{%
  \endarray \hskip -\arraycolsep
}
\newcommand{\beq}{\begin{equation}}
\newcommand{\eeq}{\end{equation}}
\newcommand{\bea}{\begin{eqnarray}}
\newcommand{\eea}{\end{eqnarray}}
\renewenvironment{pmatrix}{\left(\matrix}{\endmatrix\right)}
\newcommand{\equ}[1]{\begin{eqnarray} #1 \end{eqnarray}}

\newcommand{\ga}{\alpha}
\newcommand{\gb}{\beta}
\newcommand{\ggam}{\gamma}
\newcommand{\gd}{\delta}
\renewcommand{\ge}{\epsilon}

\newcommand{\gz}{\zeta}
\newcommand{\get}{\eta}
\newcommand{\gth}{\theta}

\newcommand{\gk}{\kappa}
\newcommand{\gl}{\lambda}
\newcommand{\gm}{\mu}
\newcommand{\gn}{\nu}
\newcommand{\gx}{\xi}
\newcommand{\gp}{\pi}

\newcommand{\gf}{\phi}
\newcommand{\gvf}{\varphi}
\newcommand{\gc}{\chi}

\newcommand{\gG}{\Gamma}
\newcommand{\gD}{\Delta}

\newcommand{\gL}{\Lambda}

\newcommand{\gP}{\Pi}

\newcommand{\gF}{\Phi}
\newcommand{\gPs}{\Psi}
\newcommand{\gO}{\Omega}
\newcommand{\cD}{{\mathcal D}}

\newcommand{\cH}{{\mathcal H}}

\newcommand{\cM}{{\mathcal M}}

\newcommand{\cQ}{{\mathcal Q}}

\newcommand{\tp}{{\tilde p}}

\newcommand{\tv}{{\tilde v}}

\newcommand{\tx}{{\tilde x}}

\newcommand{\tB}{{\tilde B}}
\newcommand{\tC}{{\tilde C}}

\newcommand{\tga}{{\tilde\alpha}}
\newcommand{\tgb}{{\tilde\beta}}
\newcommand{\tgg}{{\tilde\gamma}}

\newcommand{\tgk}{{\tilde\kappa}}

\newcommand{\tgm}{{\tilde\mu}}

\newcommand{\tgp}{{\tilde\pi}}

\newcommand{\tgvf}{{\tilde\varphi}}
\newcommand{\tgc}{{\tilde\chi}}

\newcommand{\tgPs}{{\tilde\Psi}}

\newcommand{\Bgd}{\bm{\delta}}

\newcommand{\Bgf}{\bm{\phi}}

\newcommand{\Bget}{\bm{\eta}}

\newcommand{\BgO}{\bm{\Omega}}

\newcommand{\Bgfc}{\Bgf_\chi}
\newcommand{\gvfc}{\gvf_\chi}
\newcommand{\gac}{\ga_\chi}
\newcommand{\gksq}{\kappa_0^2}
\newcommand{\cDt}{\cD_t}
\newcommand{\cDe}{\cD_\eta}
\newcommand{\Bq}{\bm{q}}
\newcommand{\Bk}{\bm{k}}

\newcommand{\geH}{\ge_\cH}
\newcommand{\getH}{\get_\cH}
\newcommand{\gthH}{\gth_\cH}
\newcommand{\zH}{z_\cH}
\newcommand{\gvfck}{\gvf_{\gc k}}

\newcommand{\mU}{\mathscr{U}}
\newcommand{\mV}{\mathscr{V}}
\newcommand{\mI}{\mathscr{I}}
\newcommand{\mJ}{\mathscr{J}}
\begin{document}

\title{Perturbative analysis of multiple-field cosmological inflation}

\author{Joydev Lahiri}
\email{joy@veccal.ernet.in}
\affiliation{Variable Energy Cyclotron Centre, 1/AF Bidhan Nagar
Kolkata 700 064, India}
\author{Gautam Bhattacharya}
\email{gautam@theory.saha.ernet.in}
\affiliation{Saha Institute of Nuclear Physics, 1/AF Bidhan Nagar
Kolkata 700 064, India}

\date{\today}

\begin{abstract}
\noindent We develop a general formalism for analyzing linear 
perturbations in multiple-field cosmological inflation based on 
the gauge-ready approach. Our inflationary model consists of
an arbitrary number of scalar fields with non-minimal kinetic terms.
We solve the equations for scalar- and tensor-type perturbations
during inflation to the first order in slow-roll, and then obtain
the super-horizon solutions for adiabatic and isocurvature perturbations 
after inflation. Analytic expressions for power-spectra and 
spectral indices arising from multiple-field inflation are presented. 
\end{abstract}


\maketitle
\section{Introduction}
\labl{intro}
Observations suggest that the early Universe underwent a period of accelarated
expansion called cosmological inflation. In addition to providing a causal
mechanism for the generation and evolution of large-scale structure formation,
inflation also leads to elegant resolutions of a number of puzzles of the 
Big-Bang theory, such as the isotropy, horizon and the flatness 
problems \cite{Guth,Lindebook,Liddlebook}. The simplest implementation of
inflation is achieved by assuming that the matter is described by a single
scalar field, the \textit{inflaton} \cite{Guth,Single}. The quantum fluctuations
of the scalar field generated during inflation become classical after crossing the
event horizon, seeding the observed density perturbations. In addition, inflation 
also generates metric perturbation in the tensor sector, leading to a stochastic
gravitational wave background. Indeed, the recent high accuracy CMBR data from
the WMAP satellite \cite{WMAP} do indeed support the general predictions of
inflationary cosmology.

It was, however, 
realized, that with a single scalar field the
inflationary scenario suffers from what is called the \textit{graceful-exit}
problem \cite{Guth,Exit}, namely,
achieving sufficient inflation consistent with the observed
density perturbations, before the Universe exits from the inflationary epoch.
Kofman et al.~\cite{Kofman} suggested the possibility
of using a first short stage of double inflation 
in order to generate a large value of
a scalar field required for a second, longer stage with a graceful exit.
Linde \cite{Linde90} showed that 
one requires at least \textit{two} scalar fields 
to overcome the graceful exit issue, without modifying
Einstein gravity, and without sacrificing natural initial conditions. 
There are other motivations for incorporating multiple fields in the 
dynamics of cosmological inflation.
For example,
when constructing models
of inflation inspired by particle physics theories such as
low energy effective supergravity derived from superstrings, one obtains
many scalar fields (see \cite{Lythrep} for a recent review). This calls for
a general framework for handling cosmological perturbations
in a situation where the matter sector consists of an arbitrary number
of scalar fields. 

Cosmological perturbations in a single field inflation has been thoroughly 
investigated in the past \cite{Liddlebook,Mukhanovetal,Kodamasasaki}.
following the seminal paper of Bardeen \cite{Bardeen80}. In the context
of multiple-field inflation, Starobinsky \cite{Starobinsky1} obtained
an expression for density perturbations with an arbitrary number of
scalar fields interacting between themselves through gravity.
In a consistent treatment of cosmological
perturbations with more than one field, one should consider the role of 
isocurvature, or entropy modes in addition to the adiabatic,
or curvature perturbations \cite{Polar1,Polar2,Polar3,Starobinsky2}. 
Indeed, it is quite possible for the 
two to be correlated, leading to distinct
observational results \cite{Lang}. 
Recently Gordon et al.~\cite{gordon} analyzed
the evolution of
adiabatic and isocurvature perturbations in  
multicomponent inflation, where
they performd a local rotation in the field space to separate out the 
adiabatic and entropy modes, while Wands et al.~\cite{wands} studied
possible observational aspects of adiabatic and isocurvature spectra
produced by two-field inflation.
For a somewhat different approach, see 
Malik et al.~\cite{Malik}. A method for treating 
density perturbations
in multicomponent inflation was proposed in \cite{Tent}, but see also
\cite{Stewart1,Stewart2,Stewart3}.

In a recent paper \cite{Lahiri1} we presented a general formalism to analyze
cosmological perturbations in inflation driven by multicomponent scalar fields
using the \textit{gauge-ready} method developed by Hwang and colleagues
\cite{Hwang1,Hwang2,Hwang3,Hwang4}. This approach follows from a suggestion by
Bardeen \cite{Bardeen88},
that rather than imposing a particular gauge condition while dealing with 
cosmological perturbations right from the beginning,
it is often advantageous to express the perturbations without specifying
any gauge. Thus one has the flexibility of adopting different gauge
conditions at a much later stage, depending upon the nature of each
problem. Moreover, it becomes easy to relate results between various
gauge-dependent and gauge-invariant techniques.

This paper is organized as follows. In Section \ref{formal} we present the
equations describing multicomponet scalar fields with a non-trivial field
metric and having non-minimal kinetic terms coupled to Einstein gravity.
By introducing a set of basis vectors based on the Gram-Schmidt 
orthonormalization technique, we can discriminate multiple-field effects
from single-field ones. Metric perturbations are discussed in Section \ref{pertuniv}.
These are conveniently decomposed into scalar, vector and tensor modes.
Here we introduce the gauge-ready approach to cosmological perturbations and 
present the equations governing density perturbations in the gauge-ready form,
as well as in terms of gauge-invariant variables. Next, we introduce the slow-roll
variables in Section \ref{solpert} and proceed to obtain the solutions to scalar
and tensor perturbations during inflation in the first order in slow-roll. We 
then proceed to study perturbations after inflation. In Section \ref{postinfl}
we discuss adiabatic and isocurvature perturbations. We calculate the 
power-spectra and spectral indices for adiabatic, isocurvature, correlated,
and tensor modes. These should be useful in comparing theoretical predictions
of various inflationary models with observations. 
We conclude in Section \ref{discuss}.

\section{The Inflationary Model}
\labl{formal}
\subsection{Scalar fields}
\labl{scalarfield}
In this Section, we explain our notation and set up the basic equations
needed for our analysis.
We consider Einstein gravity
coupled to an arbitrary
number of real scalar fields. The Lagrangian then is
\equ{
\mathscr L &=&
       \sqrt{-g} \lh \frac {1}{2 \gksq} R -
       \half \der^\gm \Bgf \cdot \der_\gm\Bgf
       - V(\Bgf) \rh                         \nonumber \\
    &=&
      \sqrt{-g} \lh \frac {1}{2 \gksq} R -
      \half g^{\mu\nu} \der_\mu \Bgf^T \mx G \der_\nu \Bgf
      - V(\Bgf)  \rh. \nonumber
\\
\labl{action}
}
were $R$ is the scalar curvature, $\gk_0^2 \equiv 8 \gp G$, and we set
$c = 1$. For the scalar fields we use a vector notation,
$\Bgf \equiv (\gf^a)$,
with the indices $a,b,c,\ldots = 1,2,3,$ $\ldots, N$
labelling the $N$--components in field space.
Further, $g \equiv {\text{det}}(g_{\gm \gn}) $,
and $\gm, \gn, \ldots$ denote the spacetime indices. For repeated indices,
the summation convention applies. The second quantity within the parentheses
of Eq.~(\ref{action}) represents a non-minimal
kinetic term. Such a kinetic term appears in various models of 
high-energy physics \cite{Lythrep}. Also $V(\Bgf)$ is an arbitrary scalar
potential.

The scalars $\Bgf$ may be interpreted as
coordinates $(\gf^a)$ on a real manifold $\cM$ induced with a symmetric
Riemannian metric $\mx G$ having components $G_{a b}$ in the
field space \cite{Tent}.
The field metric is chosen to be positive-definite so that the Hamiltonian is
bounded from below. The special case of minimally-coupled fields corresponds
to the situation $G_{a b} \equiv \gd_{a b}$.
From the components $G_{a b}$ we can define the
connection coefficients $\gG^a_{b c}$ in the usual manner,
\beq
\gG^a_{bc} = \half G^{ad}
\lh G_{bd,c} + G_{cd,b} - G_{bc,d} \rh.
\labl{connection}
\eeq
The curvature
tensor on $\cM$ is introduced in terms of the tangent vectors
$\vc{B}, \vc{C}, \vc{D}$:
\beq
[\mx{R}(\vc{B},\vc{C})\vc{D}]^a \equiv
R^a_{\; bcd} \, B^b C^c \, D^d \equiv
\lh \gG^a_{bd,c} - \gG^a_{bc,d} + \gG^e_{bd} \gG^a_{ce}
- \gG^e_{bc} \gG^a_{de} \rh  B^bC^c \, D^d .
\labl{curve}
\eeq
For any two vectors $\vc{A}$ and $\vc{B}$, we define the inner product and
the norm as
\equ{
& & \vc{A} \cdot \vc{B} =
\vc A^\dag \vc B \equiv
\vc{A}^T \mx{G} \vc{B} = A^a G_{ab} B^b , \nonumber
\\
& & |\vc{A}| \equiv \sqrt{({\vc{A}\cdot\vc{A}})} ,
\labl{prod}
}
respectively. Here $\vc A^\dag$ is the cotangent vector such that
$(\vc A^\dag)_a \equiv A^b G_{ba}$. We also introduce the covariant
derivative $\nabla_a$ on $\cM$ acting upon a vector $\vc{A}$ as
\beq
\nabla_b A^a \equiv A^a_{\; ,b} + \gG^a_{bc} A^c ,
\labl{deriv1}
\eeq
while, the covariant derivative on $\vc{A}$ with respect to
the spacetime $x^\gm$ is
\beq
\cD_\gm A^a \equiv \der_\gm + \gG^a_{bc} \der_\gm \gf^b A^c .
\labl{deriv2}
\eeq
It should be noted that the covariant derivative reduces to the
ordinary derivative when it acts upon a scalar.

By varying the action (\ref{action}) with respect to $g_{\gm \gn}$
and $\Bgf$, we obtain the gravitational field equation,
\beq
\frac {1}{\gksq} G^\gm_{\; \nu} = T^\mu_{\;\nu}
= \der^\mu \Bgf \cdot \der_\nu \Bgf
- \gd^\mu_\nu \lh \half \der^\gl \Bgf \cdot \der_\gl \Bgf + V \rh ,
\labl{tmunu}
\eeq
and the equation of motion for the scalar fields,
\beq
g^{\mu\nu} \lh \cD_\mu \gd^\gl_\nu - \gG^\gl_{\mu\nu} \rh \der_\gl \Bgf
- \mx G\inv \Bnabla^T V = 0 ,
\labl{eqmot}
\eeq
where $G^\gm_{\; \gn}$ and $T^\gm_{\; \gn}$ are Einstein and
energy-momentum tensors.

It is often convenient to represent the scalar fields
as effective fluid quantities.
The energy-momentum tensor can be covariantly decomposed into fluid
quantities using a
time-like four-vector $u^\gm$ normalized as $u^\gm u_\gm = -1$:
\equ{
& & T_{\ga \gb} = \gm u_\ga u_\gb + p h_{\ga \gb} + q_\ga u_\gb +
q_\gb u_\ga + \gp_{\ga \gb},                \nonumber
\\
& & \gm \equiv T_{\ga \gb} u^\ga u^\gb ,
\quad
p \equiv \frac{1}{3}T_{\ga \gb}h^{\ga \gb} , \nonumber
\quad
q_\ga \equiv -T_{\gb \ggam} u^\gb h^\ggam_\ga ,  \nonumber
\\
& & \gp_{\ga \gb} \equiv T_{\ggam \gd}h^\ggam_\ga h^\gd_\gb - ph_{\ga \gb}.
\labl{fluid-decomp}
}
Here $\gm$, $p$, $q_\ga$, and $\gp_{\ga \gb}$ are the energy density,
pressure, energy flux, and anisotropic pressure, respectively;
$h_{\ga \gb} \equiv g_{\ga \gb}+u_\ga u_\gb$ is a projection
tensor based on $u_\ga$ vector, $q_\ga u^\ga = 0 = \gp_{\ga \gb}$,
and $\gp^\ga_\ga = 0$.
Thus, for a multicomponent scalar field, the above decomposition gives
\equ{
& & \gm = \half|\dot{\Bgf}|^2+V ,
\quad
p = \half|\dot{\Bgf}|^2-V , \nonumber
\\
& & q_\ga = 0 = \gp_{\ga \gb}.
\labl{scalar-fluid}
}
Equations (\ref{tmunu}) and (\ref{eqmot}) with (\ref{scalar-fluid}) provide 
the fundamental expressions required for describing cosmological inflation.
\subsection{Basis vectors}
\labl{basis}
It will now prove useful to introduce a set of basis vectors
generated using Gram-Schmidt orthonormalization \cite{Tent,Malik2}.
From the vector $\Bgf$ we can construct a set of $N$ linearly
independent vectors $\{\Bgf^{(1)},\Bgf^{(2)},\ldots,\Bgf^{(N)}\}$,
where,
\equ{
\Bgf^{(1)} \equiv \dot{\Bgf},
\quad
\Bgf^{(n)} \equiv \cDt^{(n-1)}\dot{\Bgf}\;\;
\quad
(n \geq 2).
\labl{shorthand}
}
Let $\vc{e}_1 = \Bgf^{(1)}/{|\Bgf^{(1)}|}$ be the first unit vector
along the direction of the field velocity $\dot{\Bgf}$. Define the
second unit vector $\vc{e}_2$ to be along that part of the direction
of the field accelaration $\cDt \dot{\Bgf}$ which is normal
to $\vc{e}_1$ so that $\vc{e}_1 \cdot \vc{e}_2 = 0$.
Repetitively applying this Gram-Schmidt procedure
generates a set of mutually orthonormal vectors $\{\vc{e}_n\}$,
spanning the same subspace as the vectors $\{\Bgf^{(n)}\}$.

Introducing the projection operators $\mx P_n$ and $\mx P^{\perp}_n$,
which project on $\vc{e}_n$ and on the subspace perpendicular
to $\{\vc{e}_1,\ldots,\vc{e}_n\}$ respectively,
we may then write a general unit vector $\vc{e}_n$ as,
\beq
\vc{e}_n = \frac{\mx P^{\perp}_{n-1}\Bgf^{(n)}} 
{|\mx P^{\perp}_{n-1}\Bgf^{(n)}|},
\labl{en}
\eeq
where,
\equ{
\mx P_n = \vc{e}_n \vc{e}^{\dag}_n,
\quad
\mx P^{\perp}_n = \Id - \sum^{n}_{q=1}\mx P_q,
\quad
\mx P^{\perp}_0 \equiv \Id,
\labl{proj}
}
and we define,
\equ{
\mx P^{\parallel} \equiv \mx P_1 =
\vc{e}_1 \vc{e}^{\dag}_1,
\quad
\mx P^{\perp} \equiv \mx P^{\perp}_1 = \Id - \mx P^{\parallel}.
\labl{parperp1}
}
Note that when the denominator
in Eq.~(\ref{en}) vanishes, the corresponding basis vector does not
exist.

Since $\mx P^{\parallel}+\mx P^{\perp} \equiv \Id$, we can
decompose any vector $\vc{A}$ in directions parallel and perpendicular
to the field velocity:
\equ{
\vc{A} &=& \vc{A}^{\parallel} + \vc{A}^{\perp}
\equiv (\mx P^{\parallel}+ \mx P^{\perp})\vc{A} \nonumber
\\
&=& \vc{e}_1(\vc{e}_1 \cdot \vc{A})+
\vc{e}_2(\vc{e}_2 \cdot \vc{A}).
\labl{parperp2}
}
When there is just one field, $\vc{e}_1$ by definition simply
reduces to the normalized scalar $\gf^{(1)}/{|\gf^{(1)}|}$ while
$\vc{e}_2$ vanishes identically, and so do all other basis
vectors. Thus the decomposition (\ref{parperp2}) enables us to distinguish
between single-field contributions, where only $\vc{e}_1$ survives,
from multiple-field ones.

\section{The perturbed universe}
\labl{pertuniv}
\subsection{Metric perturbations}
\labl{pert}
The observed Universe is not perfectly homogeneous and
isotropic. Assuming that the inhomogeneities are small
enough, we can then treat the deviations by
considering linear perturbations of the homogeneous
and isotropic cosmological space-time described by the
Friedmann-Robertson-Walker ( FRW ) model,
\equ{
d s^2 &=& -a^2 \lh 1 + 2 A \rh d \get^2 -2 a^2 B_i d x^i \nonumber \\
      & & + a^2 \lh g^{(3)}_{i j} + 2 C_{i j} \rh d x^i d x^j ,
\labl{metric}
}
where $a(t)$ is the scale factor, $dt \equiv a d\get$,
and indices $i,j,\ldots$, run from $1$ to $3$ labelling the spatial
components. The perturbed order variables
$\;A(t, \vc{x})$, $B_i(t, \vc{x})$, and
$C_{i j}(t, \vc{x})$ are based on the metric $g^{(3)}_{i j}$
of the 3-surfaces of constant curvature $K = 0, \pm1$.
Here $t$ and
$\get$ are the comoving and conformal times respectively. We denote a
derivative with respect to comoving time by $\;\dot{} \equiv \der_t$ and
one with respect to conformal time by$\;\prime \equiv \der_\get$. The
Hubble parameters in terms of comoving and conformal times are defined
as $H = \dot a /a$ and $\cH = a' /a = aH$.

Similar to the metric decomposition Eq.~(\ref{metric}), we can
decompose the scalar field as
\beq
\Bgf(t, \vc{x})  = \bar{\Bgf}(t) + \Bgd \Bgf(t, \vc{x}) ,
\labl{phidecomp}
\eeq
where the perturbation $\Bgd\Bgf \equiv (\gd \gf^a)$
is a tangent vector on $\cM$,
while the energy-momentum tensor is decomposed as
\equ{
& & T^0_{\; 0} = - \gm \equiv -(\bar{\gm}+\gd \gm) , \nonumber
\\
& & T^0_{\; i} = \frac{1}{a}[q_i + (\gm + p)u_i]
\equiv (\gm + p)v_i , \nonumber
\\
& & T^i_{\; j} = p \gd^i_j + \gp^i_j
\equiv (\bar p + \gd p)\gd^i_j + \gp^{(3)i}_{\;\;\;\;\; j}.
\labl{tmndecomp}
}
The barred entities denote background variables. For notational
simplicity we shall ignore the overbars unless required.
In Eq.~(\ref{tmndecomp}),
$v_i$ is the frame-independent flux variable, and
$v_i$, $\gp^{(3)i}_{\;\;\;\;\; j}$ are based on
$g^{(3)}_{i j}$.

From Eqs.~(\ref{tmunu}) and
(\ref{eqmot}), the equations for the background can be written as
\equ{
& &H^2 = \frac{1}{3} \gksq  \gm - \frac{K}{a^2}
= \frac{1}{3} \gksq \lh \half |\dot  {\Bgf}|^2 + V \rh - \frac{K}{a^2} , \nonumber
\\
\labl{b1}
\\
& &\dot H = - \half \gksq \lh \gm + p \rh + \frac{K}{a^2}
= - \half \gksq \; |\dot {\Bgf}|^2 + \frac{K}{a^2} ,
\labl{b2}
\\
& &R = 6 \lh 2 H^2 + \dot H + \frac{K}{a^2} \rh ,
\labl{b3}
\\
& &\cDt \dot {\Bgf} + 3 H \dot {\Bgf} +
\mx G\inv \Bnabla^T V = 0 ,
\labl{b4}
\\
& &\dot \gm + 3 H \lh \gm + p \rh = 0.
\labl{b5}
}
We shall ignore the cosmological
constant $\gL$ in our work; nevertheless it can be easily
included by making the replacements $\gm \ra \gm + \gL/\gksq$
and $p \ra p - \gL/\gksq$. Note that we have explicitly
retained $K ( = 0, \pm 1)$, and only at a later stage shall we 
set $K = 0$.

\subsection{Scalar, vector and tensor decompositions}
\labl{svt}
To make further progress, we decompose
the perturbed order variables into scalar-, vector-, and tensor-type
perturbations. To the linear order, they
decouple from one another and evolve independently. Accordingly, the
metric perurbation variables $A(t, \vc{x})$, $B_i(t, \vc{x})$, and
$C_{i j}(t, \vc{x})$ may be decomposed as
\equ{
& & A \equiv \ga,     \nonumber
\quad
B_i \equiv \gb_i + B^{(v)}_i, \nonumber
\\
& & C_{i j} \equiv g^{(3)}_{i j}\gvf + \ggam_{, i|j} + C^{(v)}_{(i|j)}
+ C^{(t)}_{i j}.
\labl{svtdec}
}
The superscripts $(s)$,
$(v)$ and $(t)$ indicate the scalar-, vector-
and tensor-type perturbed order variables. The vertical bar represents
a covariant derivative with respect to $g^{(3)}_{i j}$ and the round
brackets in the subscript imply symmetrization of the indices. The scalar
metric perturbations are then given by $\ga$, $\gb$, $\ggam$ and $\gvf$. The
transverse-type vector perturbations $B^{(v)}_i$ and $C^{(v)}_i$ satisfy
$B^{(v) i}_{\;\;\;\;\;| i} = 0 = C^{(v) i}_{\;\;\;\;\;| i}$ 
while the tensor-type perturbation
$C^{(t)}_{i j}$ is
transverse-traceless $( C^{(t) i}_{\;\;\;\;\;i} = 0 
= C^{(t) j}_{\;\;\;\;\;i|j} )$. Both the vector and tensor perturbed order
variables are based on $g^{(3)}_{i j}$.
We define $\gD$ as a comoving three-space Laplacian, and introduce
the following combinations of the metric variables,
\equ{
& & \gc \equiv a ( \gb + a \dot{\ggam} ) , \nonumber
\quad
\gk \equiv 3 ( H \ga - \dot{\gvf} - \frac{\gD}{a^2} \gc ) , \nonumber
\\
& & \gPs^{(v)} \equiv B^{(v)} + a \dot C^{(v)}.
\labl{combi}
}

It is convenient to separate the temporal and spatial aspects of the
perturbed order variables by expanding them in terms of
harmonic eigenfunctions
$\cQ^{(s,v,t)}(\vc{k};\vc{x})$ of the generalized Helmholtz equation
\cite{Bardeen80, Kodamasasaki},
with $\vc{k}$ the wave vector in Fourier space and $k = |\vc{k}|$.
We can then write the scalar-type perturbed order variables as
$\ga(t, \vc{x}) \equiv \ga(t, \vc{k})\cQ^{(s)}(\vc{k};\vc{x})$, with
similar expressions for $\gb$, $\ggam$ and $\gvf$. The vector- and tensor-type
perturbations are expanded as $B^{(v)}_i \equiv B^{(v)}\cQ^{(v)}_i$,
$C^{(v)}_i \equiv C^{(v)}\cQ^{(v)}_i$, and
$C^{(t)}_{i j} \equiv C^{(t)}\cQ^{(t)}_{i j}$. In each of these harmonic 
expansions, a summation over the modes of the eigenfunctions is implied.
Thus, the perturbed scalar fields have the expansion
\beq
\Bgd \Bgf(t, \vc{x}) \equiv \Bgd \Bgf(t, \vc{k})\cQ^{(s)}(\vc{k};\vc{x}).
\labl{phiharmonic}
\eeq 
Similarly, the fluid variables $v_i$
and $\gp^{(3)i}_{\;\;\;\;\; j}$ can be expanded in terms of the harmonics
as
\equ{
& & v_i \equiv v^{(s)}\cQ^{(s)}_i + v^{(v)}\cQ^{(v)}_i, \nonumber
\\
& & \gp^{(3)i}_{\;\;\;\;\; j} \equiv \gp^{(s)}\cQ^{(s) i}_{\;\;\;\; j}
    + \gp^{(v)}\cQ^{(v) i}_{\;\;\;\; j}
    + \gp^{(t)}\cQ^{(t) i}_{\;\;\;\; j};
\labl{fluidharmonic}
}
while the energy-momentum tensor in Eq.~(\ref{tmndecomp}) has the expansion
\equ{
& & T^0_{\; 0} = - \gm \equiv -(\bar{\gm}+\gd \gm) , \nonumber
\\
& & T^0_{\; i} = -\frac{1}{k}(\gm + p)v^{(s)}_{, i}
    + (\gm + p)v^{(v)}\cQ^{(v)}_i ,   \nonumber
\\
& & T^i_{\; j} = (\bar p + \gd p)\gd^i_j
    + \gp^{(s)}\cQ^{(s) i}_{\;\;\;\; j}
    + \gp^{(v)}\cQ^{(v) i}_{\;\;\;\; j}
    + \gp^{(t)}\cQ^{(t) i}_{\;\;\;\; j}.
\nonumber \\
\labl{tmnharmonic}
}

For a Universe having the matter sector composed
exclusively of scalar fields, the quantity $\gp^{(3)i}_{\;\;\;\;\; j}$
in Eq.~(\ref{tmndecomp}) vanishes identically. We then have
to the perturbed order,
\equ{
& & \gd \gm = \dot{\Bgf}\cdot\cDt\gd\Bgf - \ga |\dot{\Bgf}|^2
+ \Bnabla V \cdot \gd \Bgf,
\labl{delmu}
\\
& & \gd p = \dot{\Bgf}\cdot\cDt\gd\Bgf - \ga |\dot{\Bgf}|^2 
- \Bnabla V \cdot \gd \Bgf,
\labl{delp}
\\
& & (\gm + p)v \frac{a}{k}= \dot{\Bgf} \cdot \gd \Bgf,
\labl{mup}
}
where we have written $v \equiv v^{(s)}$ for simplcity.
It is also convenient to decompose $\gd p$ into an adiabatic part
$c^2_s \gd \gm$, and an entropy perturbation $e$:
\beq
\gd p = c^2_s \gd \gm + e,
\labl{ad-ent}
\eeq
where $c^2_s \equiv \dot{p}/\dot{\gm}$ may be interpreted as an effective
sound velocity. We shall also use the notation $w \equiv p/\gm$.

\subsection{Gauge-Ready formalism}
\labl{gaugetrans}
We now briefly summarize the \textit{gauge-ready} approach discussed in
\cite{Hwang1,Hwang2,Hwang3,Hwang4,Lahiri1}. Under a \textit{gauge transformation}, or
coordinate shift, $\tx^\gm = x^\gm + \gx^\gm$
with $(\gx^0,\; \gx^i) \equiv (a\inv \gx^t,\; a \inv \gx^{, i} + \gx^{(v) i})$,
and $\gx^{(v) i}_{\;\;\;\; |i} = 0$,
the metric and matter variables transform to linear order as
\equ{
& & \tga = \ga - \dot{\gx^t}, \nonumber
\quad \tgb = \gb - \frac{1}{a}\gx^t + a \lh \frac{\gx}{a} \rh^{.},
\\
& & \tgg = \ggam - \frac{1}{a} \gx, \nonumber
\quad
\tgvf = \gvf - H\gx^t,
\quad
\tgc = \gc - \gx^t,
\\
& & \tgk = \gk + \lh 3 \dot H + \frac{\gD}{a^2} \rh, \nonumber
\quad
\tv = v - \frac {1}{a} \gx^t,
\\
& & \gd \tgm = \gd \gm - \dot \gm \gx^t, \nonumber
\quad
\gd \tp = \gd p - \dot p \gx^t,
\quad
\Bgd \tilde{\Bgf} = \Bgd \Bgf - \dot{\Bgf} \gx^t,
\\
& & \tB^{(v)}_i = B^{(v)}_i + a \dot{\gx}^{(v)}_i, \nonumber
\quad
\tC^{(v)}_i = C^{(v)}_i - \gx^{(v)}_i,
\\
& & \tv^{(v)} = v^{(v)}, \nonumber
\quad
\tgPs^{(v)} = \gPs^{(v)},
\\
& & \tgp^{(s, v, t)} = \gp^{(s, v, t)}, 
\quad
\tC^{(t)}_{i j} = C^{(t)}_{i j}.
\labl{transforms}
}
It is immediately obvious from Eq.~(\ref{transforms}) that
the tensor-type perturbations are gauge-invariant. For the
special case of scalar-type perturbations to the linear order,
fixing the temporal
part $\gx^t$ of the gauge transformation leads to different
gauge conditions: $\ga \equiv 0$ (synchronous gauge),
$\gc \equiv 0$ (zero-shear gauge), $v/k \equiv 0$ (comoving gauge),
$\gvf \equiv 0$ (uniform-curvature gauge), and so on. Except for
the synchronous gauge, the temporal gauge mode in the other gauges
are completely fixed. Consequently, there is a unique correspondence
between a variable in a gauge condition, and a gauge-invariant combination
of the variable concerned and the variable used in the gauge condition.
Using the perturbed order variables together
with the variables used in the gauge condition, one can systematically
construct various gauge-invariant variables, for example,
\equ{
& & \gvfc \equiv \gvf - H \gc, \nonumber
\quad
\gac \equiv \ga - \dot \gc,
\quad
v_\gc \equiv v - \frac{k}{a}\gc,
\\
& & \gd \gm_\gc \equiv \gd \gm - \dot{\gm} \gc,  \nonumber
\quad
\gd p_\gc \equiv \gd p - \dot{p} \gc,
\\
& & \Bgd \Bgfc \equiv \Bgd \Bgf - \dot{\Bgf}\gc, \nonumber
\quad
\Bgd \Bgf_\gvf \equiv \Bgd \Bgf - \frac{\dot{\Bgf}}{H}\gvf
\equiv - \frac{\dot{\Bgf}}{H}\gvf_{\Bgd \Bgf},
\\
& &  \gvf_v \equiv \gvf - \frac{aH}{k}v,
\quad
\gd \gm_v \equiv \gd \gm - \frac{a}{k}\dot \gm v.
\labl{givars}
}

Thus, in the zero-shear gauge, also known as the longitudinal, or
conformal Newtonian gauge, we have from Eq.~(\ref{givars}),
$\gvfc \equiv \gvf$,
$\gac \equiv \ga$, and $\Bgd \Bgfc \equiv \Bgd \Bgf$. Similarly,
in the uniform-curvature gauge, it follows that
$\Bgd \Bgf_\gvf \equiv \Bgd \Bgf$ which in turn is equivalent to
$-(\dot{\Bgf}/H)\gvf_{\Bgd \Bgf}$ in the
uniform-field gauge. In the notation of \cite{Mukhanovetal}, our
$\gac$ and $\gvfc$ correspond to their $\gF$ and $-\gPs$ respectively.

Now, as is well known in the theory
of cosmological perturbations, a judicious choice of gauge conditions
often simplifies the mathematical structure of a particular problem.
For example, density
perturbations with hydrodynamical fluids are most conveniently treated
using the comoving gauge, while gravitational potential and velocity
perturbations are best handled in the zero-shear gauge. In the same spirit,
the uniform-curvature gauge simplifies
the analysis of perturbations due to minimally coupled scalar fields.
Since, in general, we do not know the optimal gauge condition beforehand,
it becomes advantageous to express the perturbations without imposing
a specific temporal gauge condition. In other words, we write the
governing equations in the \textit{gauge-ready} form, which would give us
the freedom to choose different gauge conditions, as adapted to the
problem, at a later stage in the calculations. The equations are spatially
gauge-invariant, but the temporal gauge condition remains unspecified.
Once the temporal gauge mode
is completely fixed so that no further gauge degrees of freedom are left,
the resulting variables would then be gauge-invariant.
Moreover, when a solution in a particular gauge
is known, we can then easily derive the corresponding solution in
other gauges, as well as in gauge-invariant forms. This is the basic
concept of the gauge-ready method. The method is most useful when one
considers relativistic hydrodynamic perturbations with mutually interacting
imperfect fluids as well as kinetic components. The gauge-ready method then
not only simplifies the analysis by enabling us to choose different
gauge conditions for different aspects of the system on the fly, but also
allows us to check the numerical accuracy by comparing solutions in different
gauges.

To implement this gauge-ready strategy, it is most convenient to derive
the perturbed set of equations from the (3+1) ADM \cite{ADM}, and the
(1+3) covariant \cite{covariant} formulations of Einstein gravity. A
complete set of these equations may be found in the Appendix
of Ref.~\cite{Hwang1}. In this Section we write the equations
for scalar-type perturbations in the gauge-ready form: 

\equ{
& &\dot{\gvf}= H \ga - \frac{1}{3} \gk + \frac{1}{3} \frac{k^2}{a^2} \gc.
\labl{g1}
\\
& & - \frac{k^2-3K}{a^2} \gvf +H \gk = -\half \gksq \gd \gm.
\labl{g2}
\\
& &\gk - \frac{k^2-3K}{a^2}\gc = \frac{3}{2}\gksq(\gm+p) \frac{a}{k} v.
\labl{g3}
\\
& &\dot \gc+H\gc-\ga-\gvf=\gksq \frac{a^2}{k^2}\gp^{(s)}.
\labl{g4}
\\
& &\dot{\gk}+2H\gk+\lh 3\dot{H}-\frac{k^2}{a^2}\rh \ga=\half \gksq (\gd \gm
+3\gd p).
\labl{g5}
\\
& &\lh \cDt^2+3H\cDt-\frac{\gD}{a^2}+\mx M^2 \rh \Bgd \Bgf \nonumber
\\
& &= \lh \dot{\ga}-3\dot{\gvf}-\frac{\gD}{a^2}\gc \rh \dot{\Bgf} 
- 2\ga \mx G \inv \Bnabla ^T V.
\labl{g6}
\\
& &\gd\dot{\gm}+3H(\gd \gm+\gd p)=(\gm + p)\lh \gk-3H\ga - \frac{k}{a}v \rh.
\nonumber \\
\labl{g7}
\\
& & \frac{[a^4(\gm + p)v]^{\;\dot{}}}{a^4(\gm + p)} 
=\frac{k}{a}\left[\ga + \frac{1}{\gm + p}\lh \gd p - \frac{2}{3}
\frac{k^2-3K}{k^2}\gp^{(s)} \rh \right].
\nonumber \\
\labl{g8}
}
Equations (\ref{g1})-(\ref{g8}) are the definition of $\gk$,
the ADM energy constraint ($G^0_0$ component of the field equation),
the ADM momentum constraint ($G^0_i$ component), the
ADM propagation($G^i_j-\frac{1}{3}\gd^i_jG^k_k$ component), the
Raychaudhuri equation ($G^i_i-G^0_0$ component), the
equation of motion for scalar fields, energy conservation, and
the momentum conservation, respectively.
Here $\gd \gm$ and $\gd p$ are given by Eqs.~(\ref{delmu})
and (\ref{delp}) respectively, while
\beq
\mx M^2 = \mx G \inv \Bnabla^T \Bnabla V - \mx R (\dot{\Bgf},\dot{\Bgf}).
\labl{M-def}
\eeq
Note that these equations are valid for any $K$, and for a scalar field,
$\gp^{(s)}=0$.

Equations (\ref{g1})-(\ref{g8}), together with the background equations
(\ref{b1})-(\ref{b5}), and the perturbed order variables
for the scalar fields (\ref{delmu})-(\ref{mup}),
provide a complete set of equations for analyzing scalar-type cosmological
perturbations with multicomponent scalar
fields. As we have not chosen a specific gauge so far,
Eqs.~(\ref{g1})-(\ref{g8}) are therefore in the gauge-ready form. This
allows us to impose any one of the available temporal gauge conditions,
which would then fix the temporal gauge mode completely, leading to
gauge-invariant variables.

\subsection{Gauge-Invariant perturbation equations}
\labl{gipert}
As an illustration of the gauge-ready method, we derive some useful
expressions using the gauge-invariant variables of Eq.~(\ref{givars})
introduced in Section \ref{gaugetrans}. These may be obtained
by making judicious combinations of Eqs.~(\ref{g1})-(\ref{g8}).

Thus, from Eqs.~(\ref{g2}) and (\ref{g3}) we obtain
\beq
\frac{k^2-3K}{a^2}\gvfc = \half \gksq \gd \gm_v.
\labl{gi1}
\eeq
Eq.~(\ref{g4}) can be re-expressed as
\beq
\gac + \gvfc = -\gksq \frac{a^2}{k^2}\gp^{(s)}.
\labl{gi2}
\eeq
Eqs.~(\ref{g3}), (\ref{g4}) and (\ref{g1}) lead to
\beq
\dot{\gvf}_\gc-H\gac=-\half \gksq(\gm+p)\frac{a}{k}v_\gc.
\labl{gi3}
\eeq
Eqs.~(\ref{g7}), (\ref{g8}) with (\ref{g3}) yield
\equ{
& & \gd \dot{\gm}_v + 3H\gd \gm_v \nonumber
\\
& & = -\frac{k^2-3K}{a^2}\left[(\gm + p)\frac{a}{k}v_\gc + 
2H \frac{a^2}{k^2}\gp^{(s)}\right],
\labl{gi4}
}
while Eqs.~(\ref{g4}) and (\ref{g8}) give
\equ{
& & \dot{v}_\gc + Hv_\gc       \nonumber
\\
& & =\frac{k}{a}\left[\gac + \frac{\gd p_v}{\gm + p}-
\frac{2}{3}\frac{k^2-3K}{a^2}\frac{\gp^{(s)}}{\gm + p}\right].
\labl{gi5}
}
Combining Eqs.~(\ref{gi1})-(\ref{gi5}) we can derive
\equ{
& & \ddot{\gvf}_\gc + (4+3c^2_s)H\gvfc - c^2_s\frac{\gD}{a^2} \gvfc \nonumber
\\
& & +\left[(\gm c^2_s-p) \nonumber
-2(1+3c^2_s)\frac{K}{a^2}\right]\gvfc
\\
& & =-\half \gksq \lh e-\frac{2}{3}\gp^{(s)} \rh \nonumber
\\
& & -\half \gksq \frac{\gm +p}{H}\lh \frac{2H^2}{\gm +p}
\frac{a^2}{k^2}\gp^{(s)}\rh,
\labl{gi6}
}
where we used Eq.~(\ref{ad-ent}).
From Eq.~(\ref{gi2}) we can draw the important conclusion that, for
scalar-fields, $\gac=-\gvfc$, since $\gp^{(s)}=0$. Using this result
the equation of motion for scalar fields becomes
\equ{
\lh \cDe^2+2H\cDe-\gD +a^2\mx M^2 \rh \Bgd \Bgfc 
= -4\gvf'_\gc \Bgf'    \nonumber
\\
+ 2a^2\gvfc \mx G \inv \Bnabla ^T V,
\labl{gifield}
}
while Eqs.~(\ref{gi3}) and (\ref{gi6}) become
\equ{
\gvf'_\gc+\cH \gvfc = -\half \gksq \Bgf' \cdot \Bgd \Bgfc,
\labl{constraint}
\\
\gvf''_\gc + 6\cH \gvf'_\gc - \gD \gvfc 
+ 2\left[\cH'+2(\cH^2-K)\right]\gvfc \nonumber
\\
=\gksq a^2 \Bnabla V \cdot \Bgd \Bgfc,
\labl{gipot}
}
where we used Eq.~(\ref{mup}), and the relations
\equ{
& & e=\gd p - c^2_s\gd \gm = \gd p_\gc - c^2_s \gd \gm_\gc, \nonumber
\\
\quad
& & (1 - c^2_s)\gd \gm_\gc - e = \gd \gm_\gc-\gd p_\gc
= \Bnabla V \cdot \Bgd \Bgfc. \nonumber
\\
\labl{subsidiary}
}

Eq.~(\ref{constraint}) is often called the \textit{constraint} equation.
These equations contain most of the physics related to inflationary
cosmological perturbations. They are expressed in terms of gauge-invariant
forms of the variables, and from Section \ref{gaugetrans}
we see that they retain the same algebraic forms
in the zero-shear gauge.

We note that Eq.~(\ref{gipot}) may be recast in a different way.
According to Eq.~(\ref{parperp2}), $\Bgd \Bgfc$ may be
decomposed into components parallel and perpendicular to the field
velocity, $\Bgd \Bgfc = \Bgd \Bgf^{\parallel}_\gc + \Bgd \Bgf^{\perp}_\gc$.
Using the background equation (\ref{b4}), the constraint
equation (\ref{constraint}), and the fact 
that $|\Bgf'|'|\Bgf'| = (\cDe \Bgf') \cdot \Bgf'$, we can write
Eq.~(\ref{gipot}) as
\equ{
\gvf''_\gc &+& 2\lh \cH - \frac{|\Bgf'|'}{|\Bgf'|} \rh \gvf'_\gc \nonumber
\\
&+& 2\left[\lh \cH' - \cH \frac{|\Bgf'|'}{|\Bgf'|}\rh - 2K\right]\gvfc 
-\gD\gvfc                                                    \nonumber
\\
&=& -\gksq (\cDe \Bgf') \cdot \Bgd \Bgf^{\perp}_\gc.
\labl{newpot}
}
Following our discussion in Section \ref{basis}, we know that the
perpendicular component of field perturbation vanishes when there is
only one field. In this case, the right hand side of Eq.~(\ref{newpot})
vanishes, and the resulting equation is well known in the theory of
single field inflationary perturbations \cite{Mukhanovetal}.

\section{Perturbations during inflation}
\labl{solpert}
\subsection{Slow-roll variables}
\labl{slowroll}
Continuing with our analysis, we now
\textit{assume} that the Universe has undergone
inflation to complete flatness, so that henceforth we can set $K=0$.
This allows us to introduce the a set of functions known as
the \textit{slow-roll} variables:
\equ{
\ge(\Bgf) \equiv -\frac{\dot{H}}{H^2},
\quad
\Bget(\Bgf) \equiv \frac{\Bgf^{(2)}}{H|\dot{\Bgf}|},
\labl{slrl}
}
It is also convenient to decompose $\Bget$ into parallel and
perpendicular components using Eq.~(\ref{parperp2}),
\beq
\getp = \vc{e}_1 \cdot \Bget
= \frac{\cDt\dot{\Bgf}\cdot\dot{\Bgf}}{H|\dot{\Bgf}|^2},
\quad
\getn = \vc{e}_2 \cdot \Bget
= \frac{|(\cDt \dot{\Bgf})^\perp|}{H|\dot{\Bgf}|}.
\labl{etacomp}
\eeq

The standard slow-roll assumptions are
\beq
\ge = O(\gz),
\quad
\getp = O(\gz),
\quad
\getn = O(\gz),
\labl{slrlassump}
\eeq
for some small parameter $\gz$, with $\ge$, $\sqrt\ge\getp$ and
$\sqrt\ge\getn$ much smaller than unity.
If in an expansion in
slow-roll variables we neglect terms of order $O(\gz^2)$, we claim
that expansion to be of first order in slow-roll. Thus terms with
$\ge^2$, $\ge\getp$, etc. are of second order.
We list some useful relations involving the
slow-roll variables:
\equ{
& & \cH'=\cH^2(1-\ge), \nonumber
\quad
\frac{|\Bgf'|'}{|\Bgf'|}=\cH(1+\getp),
\\
& & \cDe \Bgf'=\cH |\Bgf'|(\Bget + \vc{e}_1)=
\gk\inv_0\sqrt{2}\cH^2\sqrt{\ge}(\Bget+\vc{e}_1), \nonumber
\\
& & \cH^2 \ge=\half \gksq |\Bgf'|^2,
\quad
\ge'=2\cH\ge(\ge+\getp).
\labl{useful}
}

\subsection{Analysis using gauge-invariant variables}
\labl{gisol}
In order to solve the system of perturbation equations (\ref{gifield}),
(\ref{constraint}) and (\ref{newpot}), we shall find it convenient to
introduce the variables,
\equ{
& &\Bq=a\lh\Bgd\Bgfc-\frac{\Bgf'}{\cH}\gvfc \rh
=a\lh\Bgd\Bgf-\frac{\Bgf'}{\cH}\gvf \rh,
\labl{sasaki1}
\\
& &u=-\frac{a}{\gksq |\Bgf'|}\gvfc
=-\frac{1}{\gk_0 \sqrt{2} H \sqrt{\ge}}\gvfc.
\labl{sasaki2}
}
Here $\Bq$ is a gauge-invariant quantity, and is a natural
generalization of the single field Sasaki-Mukhanov
variable \cite{Mukhanovetal}.

We first express the constraint equation (\ref{constraint}) in terms
of the slow-roll variables (\ref{slrl}) as
\beq
\gvfc'+\cH(1+\ge)\gvfc=-\half\gksq\Bgf'\cdot\frac{\Bq}{a}.
\labl{constq}
\eeq

The equation for the scalar field perturbations follows from (\ref{gifield}),
\beq
\cDe^2 \Bq -(\gD - \cH^2\BgO)\Bq=0,
\labl{fieldeq}
\eeq
where
\beq
\BgO = \frac{a^2\mx M^2}{\cH^2}-(2-\ge)\openone
-2\ge\lh(3+\ge)\mx P^{\parallel}
+\vc{e}_1\Bget^\dag+\Bget\vc{e}^\dag_1\rh,
\labl{omega}
\eeq
and we also used (\ref{constq}).
The corresponding Lagrangian $\mathscr L$ follows from Eq.~(\ref{fieldeq}):
\equ{
S&=&\int \mathscr L \sqrt{g^{(3)}}d\get d^3\vc{x} \nonumber
\\
&=&\half \int
\lh\cDe\Bq^\dag\cDe\Bq+\Bq^\dag(\gD-\cH^2\BgO)\Bq\rh\sqrt{g^{(3)}}
d\get d^3\vc{x}.
\nonumber \\
\labl{fieldaction}
}
Here $g^{(3)}$ is the determinant of the metric $g^{(3)}_{i j}$ of the
3-surfaces of constant curvature $K=0$, see below Eq.~(\ref{metric}).

The equation of motion for $u$ is obtained by substituting
its definition (\ref{sasaki2}) into Eq.~(\ref{newpot}):
\equ{
& & u''-\gD u-\frac{\gth''}{\gth}u =\cH \getn q_2,
\quad
q_2\equiv \vc{e}_2 \cdot \Bq, \nonumber
\\
& & \gth \equiv \frac{\cH}{a|\Bgf'|}=\frac{\gk_0}{\sqrt2}\frac{1}{a\sqrt \ge}.
\labl{ueq}
}

Further, from Eq.~(\ref{constq}) it follows that
\beq
u'+\frac{(1/\gth)'}{1/\gth}u=\half q_1,
\quad
q_1\equiv \vc{e}_1 \cdot \Bq.
\labl{constq1}
\eeq
Differentiating Eq.~(\ref{constq1}) once with respect to the conformal time
and using Eq.~(\ref{ueq}), we obtain the relation
\beq
\half \lh q'_1-\frac{(1/\gth)'}{1/\gth}q_1 \rh-\cH \getn q_2
=\gD u.
\labl{uqrel}
\eeq

We pause to note that though the equations (\ref{constq}), (\ref{fieldeq})
and (\ref{ueq}) have
been expressed in terms of the slow-roll variables, they are exact, and no
slow-roll approximation has yet been made. Observe that, to the leading
order in slow-roll, the perturbation variables $\Bq$ and $u$ decouple,
whereas at first order, mixing between these occur.

\subsection{Quantization of density perturbations}
\labl{qtmpert}
We now briefly discuss the quantization of the density perturbations described
by the Lagrangian in Eq.~(\ref{fieldaction}).
Introduce the matrix $Z_{mn}$ defined as
\beq
(Z)_{mn}=-(Z^T)_{mn}=\frac{1}{\cH}\vc{e}_m \cdot \cDe \vc{e}_n,
\labl{zmn}
\eeq
Thus $Z$ is antisymmetric and traceless. Expanding
$\Bq=q_m \vc{e}_m$ using the basis $\{\vc{e}_m\}$,
Eq.~(\ref{fieldaction}) may be expressed as,
\beq
\mathscr L=\half(q'+\cH Z q)^T(q'+\cH Z q)+\half q^T(\gD-\cH^2 \gO)q,
\labl{lagrange1}
\eeq
where $(\gO)_{mn}=\vc{e}^{\dag}_m \BgO\vc{e}_n$, and for notational ease, we
have suppresed the indices $m,\ n$. Now redefine $q$ using a new matrix $R$ as
\beq
q(\get)=R(\get)Q(\get),
\quad
R'+\cH ZR=0,
\quad
\tilde{\gO}=R^T\gO R.
\labl{qredef}
\eeq
From the equation of motion (\ref{qredef}) for $R$, it follows that $R^TR$
and $\mbox{det}R$ are constants, so that $R$ represents a rotation. Without
any loss of generality, the initial value of $R$ may be chosen as
$R(\get_0)=\openone$.
Substituting
the variables defined in Eq.~(\ref{qredef}) into Eq.~(\ref{lagrange1})
reduces the Lagrangian to the canonical form:
\beq
\mathscr L=\half Q'^T Q'+\half Q^T(\gD-\cH^2 \tilde{\gO})Q.
\labl{lagrange2}
\eeq
The corresponding Hamiltonian is then given by
\beq
\mathscr H=\half \gP^T\gP-\half Q^T(\gD-\cH^2 \tilde{\gO})Q,
\labl{hamilton}
\eeq
where the momentum $\gP$ canonically conjugate to $Q$ is
\beq
\gP(\get,\vc{x})=\partial \mathscr L /\partial Q'^T = Q'(\get,\vc{x}).
\labl{conjmtm}
\eeq

To implement the canonical quantization procedure,
the variables $(Q,\gP)$ are promoted to quantum
operators $(\hat{Q},\hat{\gP})$ satisfying the commutation relations
\equ{
& & [\ga^T\hat{Q}(\get,\vc{x}),\gb\hat{Q}(\get,\vc{x'})]=
[\ga^T\hat{\gP}(\get,\vc{x}),\gb\hat{\gP}(\get,\vc{x'})]=0, \nonumber
\\
& & [\ga^T\hat{Q}(\get,\vc{x}),\gb\hat{\gP}(\get,\vc{x'})]=
i\ga^T\gb \gd(\vc{x}-\vc{x'}),
\labl{commute1}
}
where the delta function is normalized as
\beq
\int \gd (\vc{x}-\vc{x'})\sqrt{g^{(3)}}d^3\vc{x}=1.
\labl{delta}
\eeq
Here we have introduced the vectors $\ga,\ \gb$ with components $\ga_m,\ \gb_m$
in the basis $\{\vc{e}_m\}$ to avoid writing the indices $m,\ n$ in
the commutators. Since we are considering spatially flat hypersurfaces $(K=0)$,
the operator $\hat{Q}$ may be expanded in a plane wave basis as
\beq
\hat{Q}=\int \frac {d^3\Bk}{(2\gp)^{3/2}}
\left[Q_{k}(\get)\hat{a}^\dag_{\Bk}e^{-i\Bk\cdot \mathbf{x}}+\mbox{h.c.}\right],
\labl{modexpand}
\eeq
with a similar expansion for $\hat{\gP}$. It immediately follows from
Eq.~(\ref{qredef}) that $q$ must now be interpreted as the operator
$\hat{q}$
satisfying a mode expansion identical to Eq.~(\ref{modexpand}).
The creation and annihilitation operators $\hat{a}^\dag_{\Bk}$
and $\hat{a}_{\Bk}$ satisfy
\equ{
& & [\ga^T \hat{a}_{\Bk},\gb\hat{a}_{\Bk '}]
=[\ga^T \hat{a}^\dag_{\Bk},\gb \hat{a}^\dag_{\Bk '}]=0, \nonumber
\\
& & [\ga^T \hat{a}_{\Bk},\gb \hat{a}^\dag_{\Bk '}]=\ga^T\gb \gd(\Bk-\Bk').
\labl{commute2}
}
To maintain consistency of the commutation relations (\ref{commute1}) and
(\ref{commute2}), the Wronskian condition
\beq
 W\{Q_{k},Q^*_{k}\} \equiv Q'_{k}(\get)Q^*_{k}(\get)
-Q'^*_{k}(\get)Q_{k}(\get)=i
\labl{wronski}
\eeq
must be satisfied.
From the mode expansion (\ref{modexpand}) and the
Hamiltonian (\ref{hamilton}), it follows that the equation of motion for
$Q_{k}$ is
\beq
Q''_{k}+(k^2+\cH^2\tilde{\gO})Q_{k}=0.
\labl{Qeqn}
\eeq
It may be easily verified using Eq.~(\ref{Qeqn}) that the Wronskian
satisfies $dW\{Q_{k},Q^*_{k}\}/d\get=0$.

We also interpret the variable $u$ introduced in Eq.~(\ref{sasaki2}) as an
operator $\hat{u}$, and after performing a mode expansion identical to that
of $\hat{Q}$ in Eq.~(\ref{modexpand}), it follows from Eq.~(\ref{ueq}) that
the modes $u_k$ satisfy
\beq
u''_{k}+\lh k^2-\frac{\gth''}{\gth}\rh u_k
=\cH \getn q_{2k},
\quad
q_{2k}\equiv (\vc{e}_2\cdot\vc{e}_m)q_k,
\labl{ueqk}
\eeq
or, equivalently, from Eq.~(\ref{uqrel}),
\equ{
& & \cH \getn q_{2k}-\half \lh q'_{1k}
-\frac{(1/\gth)'}{1/\gth}q_{1k} \rh
=k^2 u_k, \nonumber
\\
& & q_{1k}\equiv (\vc{e}_1\cdot\vc{e}_m)q_k.
\labl{ueqk1}
}

\subsection{First order solution}
\labl{soln}
In order to present the solution of the scalar perturbation equations,
it is convenient to introduce the time $\getH$ when the
mode with wave number $k$ crosses the Hubble radius during inflation,
so that the relation
\beq
\cH(\getH)=k
\labl{etaH}
\eeq
is satisfied for each $k$. Consequently, the inflationary epoch can be 
separated into three regions:
the \textit{sub-horizon} region ($\cH \ll k$), the \textit{transition}
region ($\cH \sim k$), and the \textit{super-horizon} region ($\cH \gg k$). We
now discuss each of these in turn.

In the sub-horizon region, we solve Eq.~(\ref{Qeqn}) with the 
$\cH^2 \tilde{\gO}$ term subdominant compared to $k^2$. The solution is obtained
in the limit $k/\cH \rightarrow \infty$ for fixed $k$ as
\beq
Q_k(\get)=\frac{1}{\sqrt{2k}}e^{ik(\get-\get_0)},
\quad
R(\get_0)=\openone.
\labl{qsubsol}
\eeq
Since one is usually interested in calculating quantities at the end of
inflation, this region is therefore irrelevant.

Considering next the transition region, we introduce
the time $\get_-$ when the sub-horizon epoch ends and the transition region
begins. In a sufficiently small interval around $\getH$ we can then 
apply slow-roll to the Eq.~(\ref{Qeqn}) keeeping all the terms, but taking
the slow-roll functions to be constant to the first order. The initial
conditions are chosen as
\beq
Q_k(\get_-)=\frac{1}{\sqrt{2k}}\openone,
\quad
Q'_k(\get_-)=\frac{i\sqrt k}{\sqrt 2}\openone,
\quad
R(\get_-)=\openone.
\labl{transinit}
\eeq
Integrating the relation for $\cH'$ in Eq.~(\ref{useful}) with the initial
conditions for the transition region, we obtain
\beq
\cH(\get)=\frac{-1}{(1-\geH)\get},
\quad
\getH=\frac{-1}{(1-\geH)k},
\labl{Hexpr}
\eeq
so that $\cH(\getH)=k$.
Differentiating $\gth$ in Eq.~(\ref{ueq}) yields
$\gth '/\gth=-\cH(1+\ge+\getp)$, which can be integrated with the result
\equ{
& & \gth(z)=\gthH \lh\frac{z}{\zH}\rh^{(1+2\geH+\get^{\parallel}_\cH)},
\quad
z\equiv k\get, \nonumber
\\
& & \gthH=\frac{\gk}{\sqrt2}\frac{H_\cH}{k\sqrt \geH},
\quad
z_\cH \equiv k\getH.
\labl{thetasol}
}
Solving Eq.~(\ref{qredef}) for the rotation matrix $R$
with the initial conditions (\ref{transinit}) yields
\beq
R(z)=\lh\frac{z}{z_-}\rh^{-(1-\geH)^{-1}Z_\cH},
\quad
z_-\equiv k\get_-.
\labl{Rsol}
\eeq
Since the time-dependent terms in the matrix $\gO$ in Eq.~(\ref{omega}) are of
first order, we can take $\gO=\gO(\getH)\equiv \gO_\cH$ in the transition 
region. Then the matrix $\tilde{\gO}$ is given by
\equ{
\tilde{\gO}&=&R^{-1}(z)\gO_{\cH} R(z)
\\
&=&\gO_{\cH}+3[\gd_{\cH},Z_{\cH}]\lh \mbox{ln}\frac{z}{z_{\cH}}+
\frac{3}{4}\mbox{ln}\;\geH\rh,
\labl{omegasol}
}
with
\equ{
\gd(\get)&=&-\frac{1}{3}\lh2\openone+\frac{\gO}{(1-\ge)^2}\rh \nonumber
\\
&=&
\ge \openone -\frac{a^2 M^2}{3\cH^2}+
2\ge (\vc{e}_1 \cdot \vc{e}_m)(\vc{e}_1 \cdot \vc{e}_n)^T, \nonumber
\\
\gd_\cH&=&\gd(\getH).
\labl{deltaH}
}
Here the second equality is valid to the first order in slow-roll, and we have
used the notation $M^2\equiv \vc{e}^{\dag}_m \mx M^2 \vc{e}_n$. We also made
the assumption that those components of $a^2 M^2/\cH^2$ which cannot
be expressed in terms of the slow-roll variables are of first order.
Because $\gd_\cH$ and $Z_\cH$ are both of first order, we can
take $\tilde{\gO}=\gO_\cH$ in Eq.~(\ref{omegasol}) to be a first
order quantity.

In order to write the equation for the mode $Q_k$ in the transition
region, we will find it convenient to define
$\bar{Q}_k \equiv R_\cH Q_k(z)$
and $\bar{\gO}=R_\cH \tilde{\gO}R^{-1}_\cH$, with $R_\cH\equiv R(z_\cH)$.
From Eq.~(\ref{Rsol}), we have to the first order, $Q_k(z)=\bar{Q}_k(z)$, while
from Eq.~(\ref{omegasol}) we conclude that $\bar{\gO}=\gO_\cH$ within a 
small region
around $z_\cH$. Using the above results in Eq.~(\ref{Qeqn}), the mode
equation for $Q_k$ may be written in terms of $\bar{Q}_k$ as
\beq
\bar{Q}_{k,\;zz}+\lh\openone -\frac{\gn^2_\cH-\frac{1}{4}}{z^2}\rh \bar{Q}_k=0,
\quad
\gn^2_\cH=\frac{9}{4}\openone+3\gd_\cH.
\labl{hankel}
\eeq
This equation is similar to the one obtained for the single-field inflation,
except that this is a matrix equation.
The solution is then given in terms of
the Hankel functions of matrix valued order $\gn_\cH$,
\equ{
& & \bar{Q}_{k}(z)=\sqrt{z}[c_1(k)H^{(1)}_{\gn_\cH}(z)
+c_2(k)H^{(2)}_{\gn_\cH}(z)], \nonumber
\\
& & \gn_\cH=\frac{3}{2}\openone+\gd_\cH.
\labl{hankelsol}
}

We wish to match the solution in Eq.~(\ref{hankelsol}) so that in the limit
$k/\cH \rightarrow \infty$, the modes approach plane waves,
$\bar{Q}_k(z)=e^{iz}/\sqrt{2k}$, see (\ref{qsubsol}). For $|z|\gg 1$, the
Hankel functions have the asymptotic forms,
\equ{
& & H^{(1)}_{\gn_\cH}(z)\sim \sqrt{2/(\gp z)}e^{i\{z-(\gn_\cH + 1/2)\gp /2\}},
\nonumber
\\
& & H^{(2)}_{\gn_\cH}(z)\sim \sqrt{2/(\gp z)}e^{-i\{z-(\gn_\cH + 1/2)\gp /2\}}.
\labl{hankelasymp}
} 
We set $c_1(k)=\sqrt{\gp/(4k)}e^{i(\gn_\cH + 1/2)\gp /2}$, and $c_2(k)=0$.
The phase factor of $c_1(k)$ is chosen in order to match with
Eq.~(\ref{qsubsol}) at short scales, while
the factor of $\sqrt{\gp/(4k)}$ ensures conformity with the Wronskian in
Eq.~(\ref{wronski}). Therefore the final solution with the appropriate
normalization is
\beq
\bar{Q}_{k}(z)=\sqrt{\gp/(4k)} e^{i(\gn_\cH + 1/2)\gp /2}
\sqrt{z}H^{(1)}_{\gn_\cH}(z).
\labl{qbarfinal}
\eeq
It is worth mentioning that that the matrix valued Hankel functions are to be
interpreted as series expansions, just like the usual Hankel functions.

We finally discuss the solution in the super-horizon region. On super-horizon
scales we have $|z|\ll 1$, for which the asymptotic form of the Hankel
function is
\beq
H^{(1)}_{\gn_\cH}(z)\sim \sqrt{2/\gp}e^{-i\gp/2}2^{\gn_\cH-3/2}
\frac{\gG(\gn_\cH)}{\gG(3/2)}z^{-\gn_\cH},
\labl{hankelsuper}
\eeq
so that the asymptotic solution for $\bar{Q}_k(z)$ in the super-horizon
region is given by
\equ{
\bar{Q}_k(z)&\sim& (1/\sqrt{2k})e^{i(\gn_\cH-1/2)\gp/2}2^{\gn_\cH-3/2}
\frac{\gG(\gn_\cH)}{\gG(3/2)}z^{\half \openone -\gn_\cH}, \nonumber
\\
&\sim&-(1/\sqrt{2k})e^{i(\gn_\cH- 1/2+2\gd_\cH)\gp /2}E_\cH
(z/z_\cH)^{-\openone - \gd_\cH },
\nonumber \\
\labl{qbarsuper}
}
where
\beq
E_\cH \equiv (1-\geH)\openone + (2-\ggam_E - \mbox{ln}\;2)\gd_\cH,
\labl{EH1}
\eeq
and $\ggam_E \approx 0.5772$ is the Euler constant.

In this region since $k/\cH \rightarrow 0$,
we can also solve Eq.~(\ref{ueqk}) ignoring the $k^2$ dependent term, 
leading to
\equ{
& &u_k(\get)=u_{P\;k}+C_k \gth
+ D_k \gth \int^{\get}_{\getH}\frac{d\get'}{\gth^{2}(\get')}, \nonumber
\\
& &u_{P\;k}=\gth \int^{\get}_{\getH}\frac{d\get'}{\gth^2}
\int^{\get'}_{\getH}d\get''\gth \cH \getn q_{2k},
\labl{ueqksol}
}
where $C_k$ and $D_k$ are constants of integration, and $u_{P\;k}$ is a
particular solution. Note that since $\gth$ is a rapidly decaying function,
we can ignore $C_k$ compared to $D_k$. In the same approximation,
the solution of Eq.~(\ref{ueqk1}) is
\beq
q_{1k}=d_{k}(1/\gth)+2(1/\gth)\int^{\get}_{\getH}d\get'\gth
\cH \getn q_{2k}.
\labl{ueqk1sol}
\eeq
From Eq.~(\ref{constq1}) we see that the integration constants $D_k$ and $d_k$
are related by $D_k=\half d_k$. Considering the region where $\get$ is
sufficiently close to $\getH$, the integral in Eq.~(\ref{ueqk1sol}) may then
be neglected, so that using Eq.~(\ref{thetasol}),
we can write $q_{1k}=2D_k (1/\gth_\cH)(z/z_\cH)^{-1}$.
Taking into account the asymptotic solution (\ref{qbarsuper}),
and the fact that $q_k=(\vc{e}_1 \cdot \vc{e}_m)^{T}q_{1k}$, we finally
obtain,
\beq
D_k=-(1/2\sqrt{2k})e^{i(\gn_\cH- 1/2+2\gd_\cH)\gp /2}
\gth_\cH (\vc{e}_1 \cdot \vc{e}_m)^{T} E_\cH.
\labl{Dksol}
\eeq
Thus the integration constant in Eq.~(\ref{ueqksol}) is
completely determined to first order in slow-roll. Inserting the result
(\ref{Dksol}) for $D_k$ in (\ref{sasaki2}) together with (\ref{ueqksol}),
and using the relation
$a_{\cH}H_{\cH}=k$, we finally arrive at
\beq
\gvfck=e^{i(\gn_\cH- 1/2+2\gd_\cH)\gp /2}\frac{\gk_0}{2k^{3/2}}
\frac{H_\cH}{\sqrt{\geH}}\left[\mI(t_{\cH},t)(\vc{e}_1
\cdot \vc{e}_m)^{T}+\mJ(t_{\cH},t) \right]E_\cH,
\labl{uksolfinal}
\eeq
where we ignored $C_k$, and
\equ{
& & \mI(t_{\cH},t)=\frac{H}{a}\int^{t}_{t_\cH}dt'a\lh\frac{1}{H}\rh^{.},
\quad
\mJ(t_{\cH},t)=\frac{H}{a}\int^{t}_{t_\cH}dt'a\lh\frac{1}{H}\rh^{.}
\mU(t_{\cH},t),                    \nonumber
\\
& & \mU(t_{\cH},t)=2\int^{t}_{t_\cH}dt'H\getn\sqrt{\frac{\geH}{\ge}}
\frac{a_\cH}{a}(\vc{e}_2\cdot \vc{e}_m)^{T}R\frac{Q_k}{Q_{k\cH}}.
\labl{ABU}
}
Here $Q_{k\cH}$ is the value of the asymptotic solution (\ref{qbarsuper}) for
$Q_k$ evaluated at $\get = \getH$. Observe that the solution
(\ref{uksolfinal}) for $\gvfck$ is expressed entirely in terms
of background quantities and comoving time.

\subsection{Vector and tensor perturbations}
For the sake of completeness, we now present a brief discussion of
vector- and tensor-type perturbations. From the $G^0_{\; i}$ component of
Eq.~(\ref{tmunu}), together with Eq.~(\ref{tmnharmonic}), we have
\beq
\half k^2 \gPs^{(v)}=\gksq a^2 (\gm + p)v^{(v)},
\labl{vectorpert1}
\eeq
while the condition $T^{\gm}_{i;\gm}=0$ yields
\beq
\frac{1}{a^4}\left[a^4 (\gm+p)v^{(v)}\right]'=-\half k\gp^{(v)}.
\labl{vectorpert2}
\eeq
Equations (\ref{vectorpert1}) and (\ref{vectorpert2}) describe the
vector-type, or rotational perturbations. 
Observe that $\gPs^{(v)}$,
$v^{(v)}$, and $\gp^{(v)}$ appearing in these equations are
gauge-invariant, see Eq.~(\ref{transforms}).
Since vector sources are
absent when the matter
sector is composed entirely of scalar fields, the vector-type perturbations
are therefore irrelevant in the inflationary scenario.

The equation for the tensor-type, or gravitational wave
perturbations follows from the $G^i_{\;j}$
component of (\ref{tmunu}):
\beq
C^{(t)\prime \prime}_{ij}+2\cH C^{(t)\prime}_{ij}
+k^2 C^{(t)}_{ij}=\gksq a^2 \gp^{(t)}_{ij}.
\labl{tensorpert1}
\eeq
For scalar fields we have $\gp^{(t)}=0$. 
Here $C^{(t)}_{ij}$ is symmetric, transverse-traceless, and gauge-invariant.
We quantize the tensor perturbations by interpreting $C^{(t)}$ as the operator
$\hat{C}^{(t)}$ with the mode expansion
\equ{
\hat{C}^{(t)}_{ij}&=&\int \frac {d^3\Bk}{(2\gp)^{3/2}}
\left[C^{(t)}_{ij\,\Bk}(\get)\hat{a}^\dag_{\Bk,\gl}e^{-i\Bk \cdot \mathbf{x}}+\mbox{h.c.}\right], \nonumber
\\
\labl{grwavedecomp1}
\\
C^{(t)}_{ij\,\Bk}(\get)&=&\sum_{\gl=+,\times}
\frac{2\gk_0}{a}\ge_{ij}(\Bk;\gl)
v_{\Bk,\gl}(\get).
\labl{grwavedecomp2}
}
The quantity $\ge_{ij}(\Bk;\gl)$ is the polarization tensor satisfying the 
conditions $\ge_{ij}=\ge_{ji}$, $\ge_{ii}=0$, $k^{i}\ge_{ij}=0$, and
$\ge^{i}_{j}(\Bk;\gl)\ge^{j*}_{i}(\Bk;\gl')=\gd_{\gl\gl'}$. The summation
is over the two independent polarization states $+$ and $\times$. As before,
the creation and annihilation operators satisfy the commutation relation,
\equ{
[\hat{a}_{\Bk,\gl},\hat{a}^\dag_{\Bk,\gl'}]=\gd_{\gl,\gl'}\gd^{(3)}(\Bk-\Bk').
\labl{gravcomm}
}
The mode $v_{\Bk,\gl}(\get)$ satisfies
\beq
v_{\Bk}'' +\lh k^2-\frac{a''}{a}\rh v_{\Bk}=0,
\labl{tensorpert}
\eeq
for each $\gl$, and $a''/a=\cH^{2}(2-\ge)$.
Proceeding similarly as in the case of scalar perturbations, the solution of
Eq.~(\ref{tensorpert}) in a sufficiently small interval of time around $\get_\cH$
to the first order in slow-roll may be written in terms of 
Hankel functions ($z=k\get$):
\equ{
v_{\Bk}(z)=\sqrt{\gp/(4k)} \sqrt{z}H^{(1)}_{\frac{3}{2}+\ge_\cH}(z).
\labl{tensorsolx}
}
Performing the asymptotic expansion of Hankel function in the region $|z|\ll1$,
the super-horizon solution for $v_{\Bk}(z)$ becomes
\equ{
v_{\Bk}(z)&\sim& (1/\sqrt{2k})e^{i(2\ge_\cH + 1)\gp/2}C_\cH(z/z_\cH)^{-1-\ge_\cH}, \nonumber
\\
C_\cH &=& 1+(1-\ggam_{E} - \mbox{ln}2)\ge_\cH.
\labl{asymptensor1}
}
On the other hand, solving (\ref{tensorpert}) in the super-horizon region yields
\beq
v_{\Bk}(z)=A_{\Bk} a+B_{\Bk} a\int^{z}\frac{1}{a^2}dz'.
\labl{tensorsoly}
\eeq
We ignore the rapidly decaying term $B_{\Bk}$, and matching the $A_{\Bk}$ term
with the solution (\ref{asymptensor1}), we finally obtain the solution of tensor
perturbation in the super-horizon region:
\beq
C^{(t)}_{ij\;\Bk}(\get)=e^{i(2\ge_\cH + 1)\gp/2}\frac{\gk_{0}\sqrt{2}}{k^{3/2}}C_{\cH}H_{\cH}
\sum_{\gl=+,\times}\ge_{ij}(\Bk;\gl)
\labl{tensorsolfinal}
\eeq

\section{Post-inflationary perturbations}
\labl{postinfl}
\subsection{Adiabatic and Isocurvature perturbations}
\labl{adiso}
In the previous Section we presented the super-horizon solutions to
scalar perturbations \textit{during} inflation. We now extend the analysis to
density perturbations \textit{after} inflation, in the regime of
radiation- / matter-domination and recombination.

It is well known that the most general density perturbation is a linear
combination of \textit{adiabatic} and \textit{isocurvature},
or entropy perturbations.
Adiabatic perturbations are just the total energy density perturbations.
They perturb the solution along the same trajectory in phase-space as
the background solution. On the other hand, isocurvature perturbations
refer to the condition when there is no perturbation in the total energy
density, there are perturbations in the \textit{ratios} of energy
densities of two or more of the components adding up to zero. Isocurvature
perturbations perturb the solution off the background solution. When there
is just one scalar field driving the inflation, isocurvature perturbations
vanish identically. However, in the multicomponent situation, not only
isocurvature perturbations are generated, but also they may be correlated with
the adiabatic perturbations \cite{bartoloetal}. According to
the definition (\ref{en}) of our basis vectors $\{\vc{e}_n\}$, when there
are $N$ scalar
fields, the adiabatic perurbation will be along the vector $\vc{e}_1$, while
the $N-1$ isocrvature perturbations will be along directions orthonormal
to $\vc{e}_1$ \cite{gordon}.

In order to simplify our analysis, we assume that, of the $N$ scalar fields,
one of them has decayed to Standard Model particles so that isocurvature
perturbations among them are absent, while the remaining $N-1$ components
have decayed to non-interacting Cold Dark Matter (CDM) \cite{langlois}. Then
the CDM components may be considered as ideal fluids without mutual
interactions.

We start with Eq.~(\ref{gi6}) in conformal time for
$K=0=\gp^{(s)}$, with the gauge-invariant entropy
perturbation $e$ defined in Eq.~(\ref{ad-ent}):
\equ{
\gvfc''&+&3\cH(1+c^2_s)\gvfc'+c^2_s k^2\gvfc
+[2\cH'+(1+3c^2_s)\cH^2]\gvfc \nonumber
\\
&=&2\gksq a^2(\gm c^2_s-p)S,
\labl{adisoeq}
}
where
\beq
S\equiv -\frac{1}{4}\frac{e}{\gm c^2_s - p} = \frac{\sum^{N-1}_{i=1} \gm_i S_i}
{\sum^{N-1}_{i=1} \gm_i},
\labl{Sdef}
\eeq
is the total entropy perturbation, and
\beq
S_i \equiv \frac{\gd \gm_i}{\gm_i + p_i}-\frac{\gd \gm_\ggam}{\gm_\ggam + p_\ggam}.
\labl{Sidef}
\eeq
In Eq.~(\ref{Sidef}) we used the notation that the Standard Model particles are
represented as photons (with subscript $\ggam$) and $p_i=0$ for CDM components,
while $p_\ggam = \gm_\ggam /3$ for photons. Since we are considering the CDM components
as mutually non-interacting ideal fluids, it can be shown that in flat space $S_i$ are constant
in the super-Hubble region \cite{Hwang5}. Further, $\gm_i$ have the same time-dependence.
Therefore, within our approximation, $S$ is constant for super-horizon scales.

\subsection{Super-Horizon solutions}
\labl{superhsol}
The solution of Eq.~(\ref{adisoeq}) in the super-horizon region, where we neglect the
$c^2_s k^2$ term, may be obtained in the same way that we derived the inflationary
solution. Introducing the variables $u$ and $\gth$ defined in Eqs.~(\ref{sasaki2})
and (\ref{ueq}) respectively, Eq.~(\ref{adisoeq}) becomes,
\beq
u''-\frac{\gth''}{\gth}u=-2\sqrt{3}\frac{a\cH}{\gk_0}\frac{c^2_s-w}{\sqrt{1+w}}S,
\labl{uadiso}
\eeq
with $w\equiv p/\gm$.
This eqution admits the solution $\gvfc=\gvfc^{(0)} + \gvfc^{(P)}$, where
\equ{
\gvfc^{(0)}&=&-\gksq C \frac{H}{a}-3D\frac{H}{a}\int^{t}_{t_0}dt'a(1+w) \nonumber
\\
&=&-\lh \gksq C + 2D\frac{a(t_0)}{H(t_0)} \rh \frac{H}{a} -2D\mI(t_0, t) \nonumber
\\
\labl{homsol}
}
is the homogeneous solution (with $S=0$), and
\equ{
\gvfc^{(P)}&=&6\frac{H}{a}\int^{t}_{t_0}dt'a(1+w)\int^{t'}_{t_0}dt''H\frac{c^2_s-w}{1+w}S \nonumber
\\
&=&2\frac{H}{a}\int^{t}_{t_0}dt'\frac{1+w}{\frac{5}{6}+\frac{1}{2}w}\lh \frac{a}{H}\rh^{.}
\int^{t'}_{t_0}dt''S \lh \frac{\frac{5}{6}+\frac{1}{2}w}{1+w}\rh^{.} \nonumber
\\
\labl{partsol}
}
is the particular solution. The function $\mI$ is defined in Eq.~(\ref{ABU}). It immediately
follows from Eq.~(\ref{partsol}) that when $S$ is constant, $\gvfc^{(P)}=2S$ is a particular
solution.

Returning back to our discussion in Section \ref{adiso}, we can claim that the adiabatic
perturbation corresponds to the homogeneous solution of Eq.~(\ref{adisoeq}) with the source
term $S$ absent, while the isocurvature perturbation pertains to the particular solution
due to the entropic source $S$. The initial conditions that one imposes on the isocurvature
perturbation are that both $\gvfc$ and $\gvfc'$ vanish at the beginning of the
radiation-dominated era. If we further make the simplifying assumption that the end of
inflation at time $t_e$ marks the immediate beginning of the radiation epoch, we can then
write the solution to isocurvature perturbation with a \textit{constant} $S$ as
\equ{
\gvfc^{(iso)}(t)&=&2S\lh 1-2 \frac{\frac{5}{6}+\frac{1}{2}w(t_e)}{1+w(t_e)}\mI(t_e, t)\rh \nonumber
\\
&=&2S\lh 1-\frac{3}{2}\mI(t_e, t) \rh.
\labl{isosol}
}
In the radiation-dominated era we have $a(t) \propto t^{1/2}$ and $w=1/3$, so that the
non-decaying part of $\mI(t_e, t)=2/3$. Thus, from Eq.~(\ref{isosol}), $\gvfc^{(iso)}$
vanishes for radiation domination. On the other hand, at the time of recombination
during matter-domination, we have
$a(t) \propto t^{2/3}$, $w=0$, and hence $\mI(t_e, t)=3/5$. In this case we have the simple
relation
\beq
\gvfc^{(iso)}=\frac{1}{5}S.
\labl{isocurvsol}
\eeq

We now derive the expression for $S$ in the super-horizon region during inflation.
Using the slow-roll functions introduced in Section \ref{slowroll}, we have
\equ{
& &\gm=\frac{1}{2}|\dot{\Bgf}|^2+V=\frac{3H^2}{\gksq}, \nonumber
\\
& & p=\frac{1}{2}|\dot{\Bgf}|^2-V=\frac{3H^2}{\gksq}\lh 1-\frac{2}{3}\ge\rh, \nonumber
\\
& &w=\frac{p}{\gm}=-1+\frac{2}{3}\ge, \nonumber
\quad
c^2_s=\frac{\dot{p}}{\dot{\gm}}=-1-\frac{2}{3}\getn,
\\
& &\gd p-c^2_s\gd \gm=\frac{2\sqrt{2}}{\gk_0}H^2\sqrt{\ge}\getn\frac{q_2}{a}
\labl{miscellany}
}
Substituting these into the definition $S=\frac{1}{4}(\gd p-c^2_s\gd \gm)/(p-c^2_s\gm)$ 
yields
\beq
S=\frac{\gk_0}{2\sqrt{2}}\frac{\sqrt{\ge}}{\ge+\getp}\getn\frac{q_2}{a}.
\labl{Sslowroll}
\eeq
From Eq.~(\ref{Sslowroll}) we see that the total entropy perturbation is generated
along the directions orthonormal to the adiabatic perturbation, in agreement with
\cite{gordon}. It is worth mentioning here that when $\dot{\ge}=2H\ge(\ge+\getp)=0$,
the expression for $S$ as given by (\ref{Sslowroll}) develops a singularity. Hence
$S$ is not a convenient variable to use during inflation. However, since we are
demanding that the transition from the end of inflation to the radiation era is
immediate, we will evaluate $S$ at the time $t_e$.

It remains to obtain the solution for the adiabatic perturbation. This is derived
by matching the solution for $\gvfc$ in Eq.~(\ref{uksolfinal}) at the end of 
inflation $t_e$ with the homogeneous solution (\ref{homsol}) while maintaining
continuity and differentiability. Ignoring the rapidly decaying
$C_k$ term, it follows that
\equ{
\gvf^{(ad)}_{\gc k}(t)&=&-2(D_k+\gth(\get_e)u_{P\;k}'(\get_e)
-\gth'(\get_e)u_{P\;k}(\get_e)) \nonumber
\\
&=&e^{i(\gn_\cH- 1/2+2\gd_\cH)\gp /2}\frac{\gk_0}{2k^{3/2}}
\frac{H_\cH}{\sqrt{\geH}}\mI(t_e,t)\left[
(\vc{e}_1\cdot \vc{e}_m)^{T} + \mU(t_e,t) \right]E_\cH. \nonumber
\\
\labl{adiabsol}
}
The functions $\mI$ and $\mU$ are defined in Eq.~(\ref{ABU}).

\subsection{Power spectra and spectral indices from inflation}
\labl{spectra}
Having obtained the solutions for adiabatic and isocurvature perturbations
given by Eqs.~(\ref{adiabsol}) and (\ref{isocurvsol}) respectively, and the
solution for the tensor perturbation (\ref{tensorsolfinal}), we are now in a
position to calculate the power spectra and spectral indices from inflation.
The power spectra are conventionally defined as
\equ{
& &\gD^{2}_{\mathcal A}(k)\equiv \frac{k^3}{2\gp^2}\langle|\gvf^{(ad)}_{\gc k}|^2\rangle,
\quad
\gD^{2}_{\mathcal S}(k)\equiv \frac{k^3}{2\gp^2}\langle|\gvf^{(iso)}_{\gc k}|^2\rangle, \nonumber
\\
& &\gD^{2}_{T}(k)\equiv \frac{k^3}{2\gp^2}\langle|C^{(t)}_{ij\,\Bk}C^{(t)\;ij}_{\Bk}|\rangle, \nonumber
\\
& &\gD^{2}_{\mathcal C}(k)\equiv \frac{k^3}{2\gp^2}\lh\langle|\gvf^{(ad)}_{\gc k}\gvf^{(iso)}_{\gc k}|
+|\gvf^{(iso)}_{\gc k}\gvf^{(ad)}_{\gc k}|\rangle\rh.
\labl{spectradef}
}
The subscripts $\mathcal{A},\ \mathcal{S}$, and $T$ refer to the adiabatic, isocurvature,
and tensor power spectra, while
$\gD^{2}_{\mathcal C}(k)$ is the spectrum due to cross-correlation between adiabatic and isocurvature
perturbations. Substituting the results for $\gvf^{(ad)}_{\gc k}$, $\gvf^{(iso)}_{\gc k}$, 
and $C^{(t)}_{ij\,\Bk}$ in the above expressions, and the using the fact that 
$\mI(t_e,t)=3/5$ during matter-domination, we obtain,
\equ{
\gD^{2}_{\mathcal A}(k)&=&\frac{9}{25}\frac{H^{2}_{\cH}}{\gp M^{2}_{Pl}\ge_{\cH}}\left[
(1-2\ge_{\cH})(1+\mU^{T}_{e}\mU_{e})-2C_{0}\lh (2\ge_{\cH}+\get^{\parallel}_{\cH})
+2\get^{\perp}_{\cH} (\vc{e}_2\cdot \vc{e}_m)^{T}\mU_{e}
+\mU^{T}_{e}\gd_{\cH}\mU_{e}\rh \right],
\labl{spectrsol1}
\\
\gD^{2}_{\mathcal S}(k)&=&\frac{1}{25}\frac{H^{2}_{\cH}}{\gp M^{2}_{Pl}\ge_{\cH}}\left[
(1-2\ge_{\cH})\mV^{T}_{e}\mV_{e}-2C_{0}\mV^{T}_{e}\gd_{\cH}\mV_{e} \right],
\labl{spectrsol2}
\\
\gD^{2}_{\mathcal C}(k)&=&\frac{3}{25}\frac{H^{2}_{\cH}}{\gp M^{2}_{Pl}\ge_{\cH}}\left[
(1-2\ge_{\cH})\mV^{T}_{e}\mU_{e}-2C_{0}\lh \get^{\perp}_{\cH}(\vc{e}_2\cdot \vc{e}_m)^{T}\mV_{e}
+ \mV^{T}_{e}\gd_{\cH}\mU_{e}\rh \right],
\labl{spectrsol3}
\\
\gD^{2}_{T}(k)&=&16\frac{H^{2}_{\cH}}{\gp M^{2}_{Pl}}\left[1-2(C_{0}+1)\ge_{\cH}) \right],
\labl{spectrsol4}
}
where $M_{Pl}=G^{-1/2}$ is the Planck Mass, and 
\beq
\mV=\frac{1}{2}\sqrt{\ge_\cH}\frac{\sqrt{\ge}\getn}{\ge+\getp}\frac{Q}{Q_\cH}
(\vc{e}_2\cdot \vc{e}_m)^{T},
\quad
C_0=\ggam_{E}+\mbox{ln}2-2.
\labl{Vexp}
\eeq
The subscript $e$ reminds us that $\mU$ and $\mV$ are evaluated at $t=t_e$. 

If we now assume that the power spectra $\gD^{2}_{X}(k)$ depend weakly on $k$, 
where $X$ denotes adiabatic, isocurvature or cross-correlated spectra,
we can parametrize them as
\beq
\gD^{2}_{X}(k)=\gD^{2}_{X}(k_0)\lh \frac{k}{k_0} \rh^{n_{X}(k_{0})-1},
\labl{pspectra1}
\eeq
while the tensor spectrum is conventionally parametrized as
\beq
\gD^{2}_{T}(k)=\gD^{2}_{T}(k_0)\lh \frac{k}{k_0} \rh^{n_{T}(k_{0})}.
\labl{tenspectra}
\eeq
The normalization factor $\gD^{2}(k_0)$ is called the amplitude, and 
$n$ the spectral index. Here $k_0$ is a suitable pivot wavenumber. These parametrizations
are valid for a range of $k$ when $n_{X}-1$ and $n_{T}$ are close to zero,
that is, when the power spectra are near \textit{scale-invariant}. They lead to the
definition of the spectral indices,
\beq
n_{X}(k_0)-1=\frac{d\;\mbox{ln}\gD^{2}_{X}(k)}{d\;\mbox{ln}k},
\labl{specdefscalar}
\eeq
for the scalar modes
and
\beq
n_{T}(k_0)=\frac{d\;\mbox{ln}\gD^{2}_{T}(k)}{d\;\mbox{ln}k},
\labl{specdeftensor}
\eeq
for the tensor modes, with the right hand side of Eqs.~(\ref{specdefscalar}) and 
\ref{specdeftensor} are to be evaluated at $k=k_0$. Substituting the values of
$\gD^{2}(k)$ from above leads to the expressions for the spectral indices
valid to first order in slow-roll:
\equ{
n_{\mathcal A}-1&=&-4\ge_{\cH}-2\get^{\parallel}_{\cH}
+\frac{2\mU^{T}_{e}(2\ge_{\cH}+\get^{\parallel}_{\cH}-\gd_{\cH})\mU_{e}-
4\get^{\perp}_{\cH}(\vc{e}_2\cdot \vc{e}_m)^{T}\mU_{e}}{1+\mU^{T}_{e}\mU_{e}},
\\
n_{\mathcal S}-1&=&-2\frac{\mV^{T}_{e}\gd_{\cH}\mV_{e}}{\mV^{T}_{e}\mV_{e}},
\\
n_{\mathcal C}-1&=&-2\frac{\mV^{T}_{e}\gd_{\cH}\mU_{e}
+\get^{\perp}_{\cH}(\vc{e}_2\cdot \vc{e}_m)^{T}\mV_{e}}{\mV^{T}_{e}\mU_{e}},
\\
n_{T}&=&-2\ge_{\cH}+\left[-2\ge^{2}_{\cH}-4(C_{0}+1)\ge_{\cH}(\ge_{\cH}
+\get^{\parallel}_{\cH} ) \right].
\labl{specindices}
}
The contribution due to multicomponet scalar fields come from $\mU$ and
$\mV$. In the case of a single scalar field, both these terms vanish, and we
recover the single-field results. Observe that in this case $n_{\mathcal S}$ and 
$n_{\mathcal C}$ are irrelevant. 

We would like to point out here that one often comes across the adiabatic power
spectrum defined in terms of the curvature perturbation 
$\mathcal{R}= -\gvf_{\gc}+(H/\dot{H})(\dot{\gvf_{\gc}}+H\gvf_{\gc})$ and denoted
by $\gD^{2}_{\mathcal R}$ \cite{gordon,wands,WMAP2}. 
This definition simply removes the time-dependent
factor $\mI$ in Eq.~(\ref{homsol}), so that
$\gD^{2}_{\mathcal A}=(9/25)\gD^{2}_{\mathcal R}$, and $n_{\mathcal A}=n_{\mathcal R}$.

Using the above results we obtain the 
\textit{consistency relation}, or the tensor to scalar ratio \cite{Lidsey},
\equ{
r&=&\frac{\gD^{2}_{T}(k_{0})}{\gD^{2}_{\mathcal R}(k_{0})} \nonumber
\\
&\simeq &16\ge_{\cH}\lh 1+\mU^{T}_{e}\mU_{e} \rh^{-1}
\simeq -8n_{T}\lh 1+\mU^{T}_{e}\mU_{e}\rh^{-1}. \nonumber
\\
\labl{consistrel}
}
The single-field result $r\simeq 16\ge_{\cH}\simeq -8n_{T}$ follows immediately from above,
while in the case of two fields, Eq.~(\ref{consistrel}) may be 
written as \cite{bartoloetal,wands},
\beq
r \simeq -8n_{T}\mbox{sin}^{2}\gD,
\quad
\mbox{cos}\gD=\gD^{2}_{\mathcal C}/{\sqrt{\gD^{2}_{\mathcal A}\gD^{2}_{\mathcal S}}}.
\labl{twoconsist}
\eeq
For single-field inflation, WMAP limits the tensor to scalar ratio as
$r(k_{0}=0.002\ \mbox{Mpc}^{-1}) < 1.28\ (95 \%\ \mbox{CL})$ \cite{WMAP2}.
An independent estimation of $n_T$ and $r$ would provide a decisive test for 
discriminating multiple-field inflation from single-field slow-roll 
models \cite{Polarski-consist}. The results obtained in this
Section may be applied towards identifying classes of inflation models
differing by their observational signatures \cite{Leach}. They may
also be used to obtain information about the slow-roll potential by 
an inverse analysis of the observational results \cite{Lidsey}.
\section{Conclusion}
\labl{discuss}
The foregoing Sections contain a general framework for analyzing the dynamics
of cosmological perturbations in multiple-field inflation using the gauge-ready
approach. The model comprises arbitrary number of scalar fields induced with a 
general field metric coupled to Einstein gravity. The complete set of equations
describing scalar perturbations were presented in the gauge-ready form as well as 
in terms of gauge-invariant variables. Solutions for density perturbations
were derived to the first order in slow roll during inflation, and for adiabatic
and isocurvature modes after inflation. Tensor perturbations were also discussed.
Expressions of the power-spectra and spectral indices for adiabatic, isocurvature,
cross-correlated and tensor modes were obtained within the first order slow roll
approximation. These results are of direct relevance when comparing theoretical
predictions of various inflation models with observations. It would be of interest
to interface our approach with the CMBFAST \cite{cmbfast} or CAMB \cite{camb} 
computer codes and compare with the WMAP results.


\end{document}